\documentclass[a4paper,11pt]{article}
\usepackage{jheppub} 
\usepackage{lineno}
\usepackage{float}
\usepackage{hyperref}

\usepackage{xcolor} 
\usepackage{soul}   
\definecolor{myorange}{RGB}{255,165,0}

\sethlcolor{yellow} 
\definecolor{mylightgreen}{RGB}{144,238,144}  

\sethlcolor{yellow} 

\title{PCN: A Deep Learning Approach to Jet Tagging Utilizing Novel Graph Construction Methods and Chebyshev Graph Convolutions}

\author[a,1]{Yash Semlani}
\author[b,1]{Mihir Relan}
\author[c,2]{Krithik Ramesh}
\affiliation[a]{University of North Carolina at Chapel Hill,\\
Chapel Hill, NC, USA}
\affiliation[b]{Johns Hopkins University,\\
3400 N. Charles St, Baltimore, MD, USA}
\affiliation[c]{Massachusetts Institute of Technology \\ 32 Vassar St, Cambridge, MA}

\emailAdd{yvsemlani@unc.edu, mrelan1@jh.edu, krithik@mit.edu}

\abstract{Jet tagging is a classification problem in high-energy physics experiments that aims to identify the collimated sprays of subatomic particles, jets, from particle collisions and ‘tag’ them to their emitter particle. Advances in jet tagging present opportunities for searches of new physics beyond the Standard Model. Current approaches use deep learning to uncover hidden patterns in complex collision data. However, the representation of jets as inputs to a deep learning model have been varied, and often, informative features are withheld from models. In this study, we propose a graph-based representation of a jet that encodes the most information possible. To learn best from this representation, we design Particle Chebyshev Network (PCN), a graph neural network (GNN) using Chebyshev graph convolutions (ChebConv). ChebConv has been demonstrated as an effective alternative to classical graph convolutions in GNNs and has yet to be explored in jet tagging. PCN achieves a substantial improvement in accuracy over existing taggers and opens the door to future studies into graph-based representations of jets and ChebConv layers in high-energy physics experiments. Code is available at \url{https://github.com/YVSemlani/PCN-Jet-Tagging}}


\begin{document}
\maketitle
\flushbottom

\section{Introduction}
The Large Hadron Collider (LHC) at CERN is a high-energy particle accelerator that collides particles to detect novel physics findings, such as the Higgs boson discovery \cite{1,2}. Proton-proton collisions will continue to advance research beyond the Standard Model, particularly in higher luminosity phases, which will provide more data to be gathered for observation. Machine learning is a pragmatic approach to analyzing the data from collider physics. One of its primary applications is jet tagging. Jet tagging is the process of identifying the elementary particle responsible for initiating a jet, which is the fragmentation and hadronization of partons. Ongoing research in jet tagging is crucial to studying the fundamental properties and interactions of elementary particles and the search for physics beyond the Standard Model. Jets are difficult to classify because machine learning models must effectively separate background jets from signal jets, requiring a generalized understanding of Quantum Chromodynamics (QCD).

A prevailing challenge in the field is developing an expressive representation of jets that captures complex relational information between jets. A variety of strategies for representing jets have been explored, including particle clouds \cite{3,4,5,6,7,8,9,10,11,12,13,14,15,16,17,18,19}, ordered sequences, and images \cite{20,21,22,23,24,25,26,27,28,29,30,31,32, 33,34,35,36,37,38}. Notably, particle cloud and graph representations have seen widespread success in conjunction with transformer architectures and graph neural networks (GNNs), consistently surpassing the performance of non-geometric methods like convolutional neural networks (CNNs), recurrent neural networks (RNNs), and multilayer perceptrons MLPs). This advantage underscores the importance of a representation that aligns with the fundamental nature of jets. Particle clouds, in particular, are well-suited due to their preservation of permutation, rotational, and translational invariance properties.  However, a limitation of particle clouds is their lack of explicit modeling of inter-particle relations within the jet structure, which contributes significantly to the fragmentation processes that take place during jet formation.

The choice of jet representation significantly influences the capabilities of machine learning models in this domain. The state-of-the-art employs particle clouds often augmented with particle interaction information.  Graph-based representations present a compelling alternative, with nodes representing particles and edges encoding relational features.  It is worth noting that though graphs can theoretically encompass the same information as particle clouds, current graph-based approaches often fall short.

Differences in model architecture can present significant opportunities for increasing model performance when combined with a suitable representation. Qu. et al. \cite{3} utilizes a transformer architecture that incorporates relational information into prediction through the particle attention layer. However, transformer models lack methods for analyzing these relational features separately. In contrast, graph-based architectures utilize edge convolutions and graph convolutions, which are heavily influenced by the relational structure of the graph. Our main contribution is the development of a graph-based representation of jets that incorporates comprehensive particle-level features and the usage of Chebyshev graph convolutions to synthesize information across disparate spatial scales.

\section{Related Works}

Deep learning using GNNs is a natural choice for jet tagging as jets are most naturally represented using graph-based representations, such as particle clouds. This makes GNNs particularly well-suited for capturing the relational information within jets and extracting features for accurate tagging. Studies using GNNs for jet tagging have explored different graph construction methods for jet representation and model architectures for optimal performance. 

The usage of a particle cloud jet representation was first proposed by Qu \& Gouskos \cite{5}, who built upon the notion of graph construction based on point-cloud structures \cite{39} and treat a jet as an invariant set of particles plotted in the $\eta$-$\phi$ space. Gong et al. \cite{6} advanced these particle cloud representations by making the edges of their graph represent particle interactions in the Minkowski space. Qu et al. \cite{3} construct their particle clouds in the $\eta$-$\phi$ space with each node representing a particle and edges representing particle interactions. Another graph-based representation by Dreyer et al. \cite{7} introduced using the Lund tree of a jet, which encodes a jet’s clustering history and substructure. In these graph construction methods, kinematic, identification, and trajectory properties are provided as node features. Dreyer et al. \cite{7} provide 5 features per node, Gong et al. \cite{6} provide 12 features per node, and Qu et al. \cite{3} provide 16 features per node. Models with more information encoded performed experimentally better, with Qu et al. \cite{3} demonstrating the best performance, followed by Gong et al. \cite{6} and Dreyer et al. \cite{7} respectively. 

Model architectures yield very different performance on classification accuracy in jet tagging. ParticleNet by Qu \& Gouskos \cite{5} utilizes Edge Convolution (EdgeConv) blocks to classify using the particle cloud representation. LundNet by Dreyer et al. \cite{7} follows a similar model architecture using EdgeConv blocks. LorentzNet by Gong et al. \cite{6} utilizes Lorentz Equivariance Group Blocks (LGEB) consisting of multiple multilayer perceptrons (MLPs) and encoder-decoder layers. Particle Transformer (ParT) by Qu et al. \cite{3} utilizes attention blocks consisting of multi-head attention modules. These mentioned model architectures used techniques that catered to the particle cloud representation and their respective graph construction methods. For a graph-based representation, however, graph convolutional layers to extract relational information would be useful to complement EdgeConv blocks. A notable advance in GNNs is the Chebyshev graph convolution (ChebConv) \cite{40}, which primarily prevents the explosion of the graph laplacian when raised to the i$^{th}$ power in a given convolutional filter. The usage of ChebConv in graph convolutional networks was shown to be effective in different tasks \cite{40,41,42}, and its usage in jet tagging has yet to be explored.

\section{Methodology}

Our jet tagging approach integrates salient features from existing methodologies, most notably ParT, Chebyshev graph convolutions, and other prominent graph neural networks. The subsequent sections will describe the core components of our study.

\subsection{Graph Construction}

A jet $j$ is a collimated spray of subatomic particles from a collision. We define $j$ by its constituent particles $j_i = $ \{$x_1, ..., x_n$\}, where $n$ is the number of particles in the jet. Note that the number of particles in each jet may be different. Each particle $x$ within a jet $j$ can be represented as a vector of features $x_i = $ \{$f_1, ..., f_{16}$\}, where $f$ is one of the 16 features provided in the dataset.

We represent each jet as a graph denoted as $G=(V,E)$, where $V$ is the set of vertices and $E$ is the set of edges. We define the vertices $V$ as \{1, ..., $n$\}, where each particle within the jet is considered to be a vertex. We define the edges as $E=\{(u,v)|u,v \in V\}$, where u and v are vertices defined by a k-Nearest Neighbors (kNN) algorithm.

It is imperative to determine the appropriate value for $k$ (number of nearest neighbors). Selections of $k$ that are excessively low result in denser graphs, fostering increased interconnectivity among particles and capturing more localized interactions but introducing noise. Conversely, excessively high values of $k$ yield sparser graphs with fewer interconnections between particles, potentially capturing more global features but overlooking crucial local interactions. Moreover, higher values for $k$ result in the exclusion of smaller jets, given the constraint that $k$ must be less than the number of particles in a jet. Notably, when $k$ is set within range of 7-10 (the ideal value suggested by the elbow method), a proportion of the \textit{H$\rightarrow$lvqq'} jets is omitted from the training dataset. Preserving the representation of these smaller jets ensures the model's proficiency in discerning subtle interactions within collisions. Utilizing the elbow method while seeking to prevent the omission of jets, we find the optimal value of $k$ as $k$ = 3. 

Our jet, $j$, is given as input to the Scipy cKDTree which creates a binary tree for quick lookup of nearest neighbors using a dimensional heuristic further explained by Maneewongvatana et al. \cite{43}. Each node's three nearest neighbors are then retrieved and linked together with an edge. Finally, the completed graph is outputted as an adjacency matrix for use in training. Figure 1 visualizes the graph-based representation versus a standard particle cloud.

Our pre-processing method differs most significantly from other graph-based representations in the features provided at each node. LundNet \cite{7}, a highly accurate graph-based representation, provides fewer features per node than our pre-processing method, meaning we encode more information overall. The features provided to the model by the authors of LundNet are also high-level features that can be derived from kinematics, which can hinder the creation of new, non-theory-based features by a model. Investigations of LorentzNet \cite{6} and ParticleNet \cite{5} follow very similar pre-processing methods to ours. In relation to these studies, our major difference is in the model architecture we utilize for classification.

\begin{figure}[t]
\centering
\includegraphics[width=13cm]{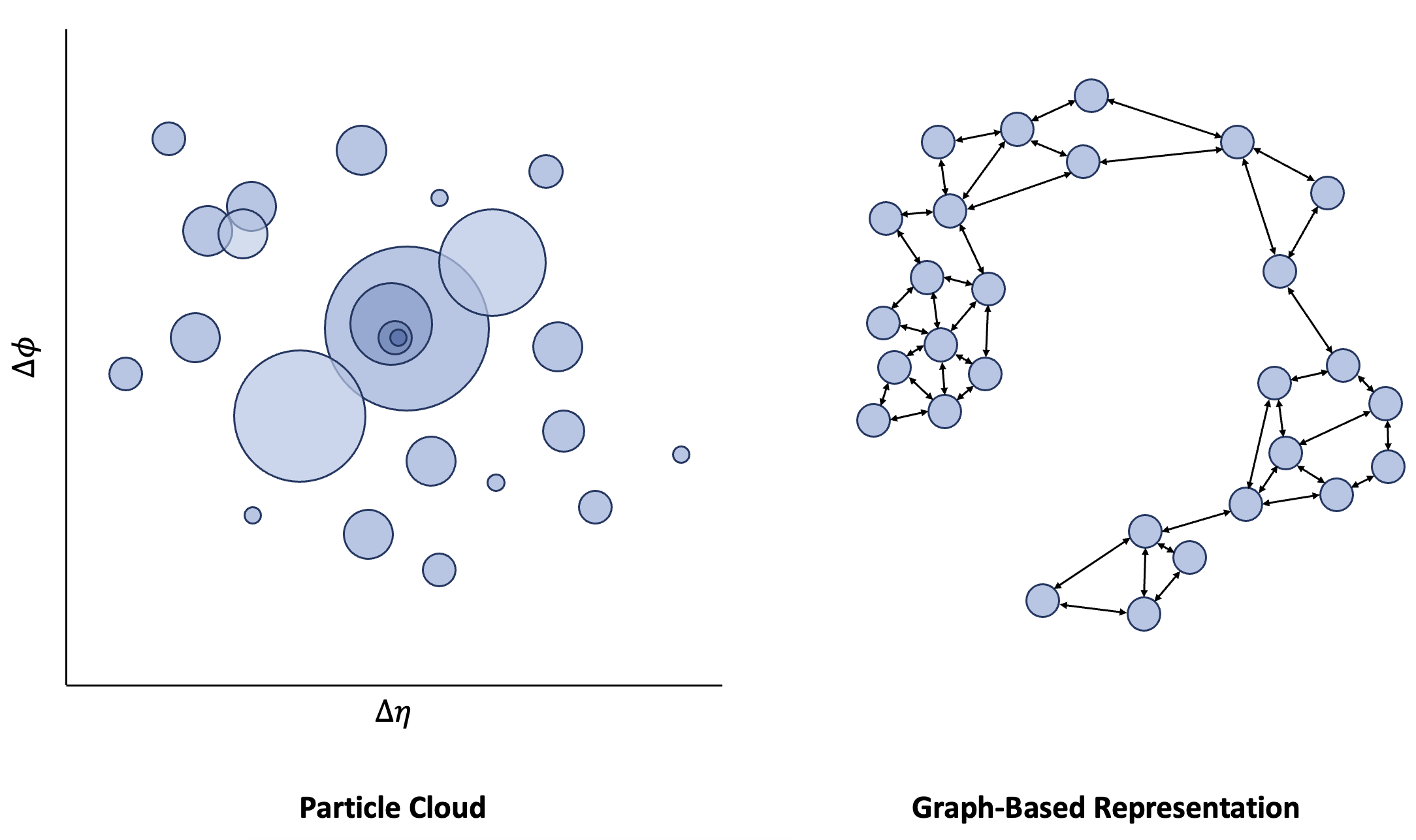}
\caption{(\textit{left}): particle cloud representation of a jet used by state-of-the-art ParT \cite{3}. (\textit{right}): graph-based representation of a jet used by PCN and PCN-Lite.\label{fig:1}}
\end{figure}

\subsection{Model Architectures}

Our models, Particle Chebyshev Network (PCN) and the streamlined PCN-Lite, leverage Chebyshev Convolutional layers (ChebConv) to process the constructed graphs (See Section 3.1). PCN  incorporates both ChebConv and Edge Convolutional (EdgeConv) layers for enhanced feature extraction while PCN-lite uses only EdgeConv. Figure 2 illustrates the model architectures.

\subsubsection{Chebyshev Graph Convolutions for Localized Interaction Analysis}

The first stage of PCN involves processing the graph-based representation of a particle jet, expressed as \( \boldsymbol{G} \in \mathbb{R}^{N \times d} \), where \( N \) represents the number of particles (nodes) in the jet and \( d \) is the dimensionality of node features. In each layer \(\ell\), the graph \( \boldsymbol{G} \) undergoes transformation through Chebyshev graph convolutions, focusing on local particle interactions. The convolution applies a series of learnable filters across the graph's nodes, considering their local neighborhood, defined by the graph structure. This process, crucial for discerning subtle particle dynamics, enables the model to capture local dependencies intrinsic to jet formation. Concurrently, these layers employ Chebyshev polynomials to adaptively filter node features, ensuring the capture of relevant local information within the jet.

\begin{figure}[t]
\centering
\includegraphics[width=13cm]{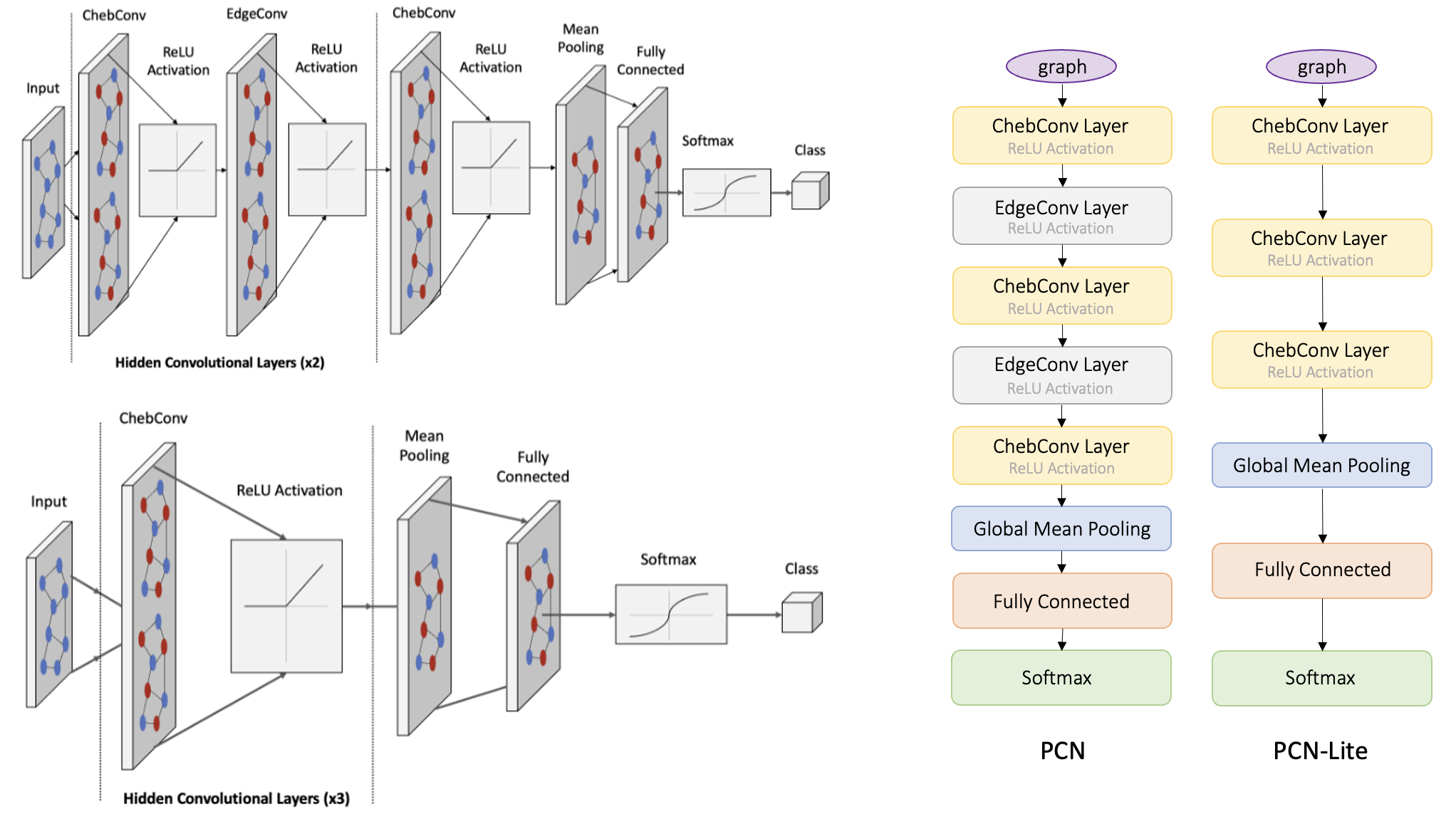}
\caption{(\textit{left}): graph neural network classification pathway of input graphs to PCN and PCN-Lite. (\textit{right}): the network architectures of PCN and PCN-Lite.\label{fig:2}}
\end{figure}

The Chebyshev convolutional operation can be formulated as:
\begin{align}
\boldsymbol{G'}_{\text{Cheb}} = \text{ChebConv}\left(\boldsymbol{G}, \boldsymbol{W}_{\text{Cheb}}^{\ell}\right)
\end{align}
where \(\boldsymbol{W}_{\text{Cheb}}^{\ell}\) represents the learnable weights in the Chebyshev convolution at layer \(\ell\).

Consider the 16-dimensional particle cloud consisting of N particles, which is then converted into a graph using the method explained in Section 3.1. A general graph convolutional operator is defined as $x' = p_{w}(L)x$, where $x$ is the feature vector of the graph and $p_{w}(L)$ is a polynomial of form:

\begin{equation} \label{eq:1}
p_{w}(L) = \sum_{i=0}^d w_{i}L^i
\end{equation}

\noindent where $w$ is a learnable parameter representing the weight of the polynomial term. The weights are learned during training to adapt the operator to the graph data. For any given feature vector $x$, it is convolved with its neighboring nodes such that the fully expanded operator is:

\begin{equation} \label{eq:2}
x'_{v} = \sum_{i=0}^{d} w_{i}\sum_{u\in N(v)}L_{u,v}^i x_{u}
\end{equation}

\noindent where $N(v)$ is the neighboring nodes of the convolved $x_{v}$. The Chebyshev convolutional operator differs from this by passing the normalized graph laplacian through a Chebyshev polynomial to create filters of the form: 

\begin{equation} \label{eq:3}
p_{w}(L) = \sum_{i=0}^d w_{i}T_{i}(L_{norm})
\end{equation}

\noindent Therefore, the full Chebyshev convolutional operator is:

\begin{equation} \label{eq:4}
x' = \sum_{i=0}^{d} w_{i}T_{i}(L_{norm})x
\end{equation}

\noindent and can be expanded as:

\begin{equation} \label{eq:5}
x'_{v} = \sum_{i=0}^{d} w_{i}\sum_{u\in N(v)}T_{i}(L_{norm})_{u,v}x.
\end{equation}

$T_{i}$ is the $i^{th}$ Chebyshev polynomial computed with initial conditions $T_{0}(x) = 1, T_{1}(x) = x$ and definition:

\begin{equation} \label{eq:6}
T_{i}(x) = 2xT_{i-1}-T_{i-2}x.
\end{equation}

\noindent $L_{norm}$ is the normalized graph Laplacian defined as:

\begin{equation} \label{eq:7}
L_{norm} = \frac{2L}{\lambda_{max}} - I
\end{equation}

\noindent where $\lambda_{max}$ is the largest eigenvalue of $L$ and $I$ is the identity matrix. The Chebyshev polynomial offers computational efficiency by approximating certain graph operations without fully diagonalizing the Laplacian, which becomes infeasible at large scales of data. 

The local neighborhood $N(v)$ over which the convolution spans is decided by the $K$ polynomial. The $K$ polynomial, $K_{k}(x)$, is defined as the difference between the $k^{th}$ Chebyshev polynomial, $T_{k}(x)$, and the $(k-2)^{th}$ Chebyshev polynomial, $T_{k-2}(x)$. By analyzing the skipping distance in graphs of $K_{k}(x)$ compared to $T_{k}(x)$, we gain valuable insights into the polynomial's approximation properties, which directly influence the performance of Chebyshev convolutions. Understanding how the number of intersections with the x-axis between consecutive zeros differs between $T_{k}(x)$ and $K_{k}(x)$ allows us to refine the techniques used for particle identification and classification in jet tagging applications. For our models, we set the $k$-value of the Chebyshev polynomial kernel to two, such that it is one less than the number of nearest neighbors per nearest neighbor node retrieval to capture substructure interactions.

This convolutional method is similar to image convolutions in utilizing $x(v)$ as analogous to a central pixel in image convolution and $N(v)$, the neighborhood of nodes around $x(v)$, as analogous to a patch in image convolution. We found that ChebConv yielded better performance compared to classical graphical convolutions. With classical graphical convolutions, a sum of neighbor node features is multiplied by a learnable weight matrix, which may lead to oversmoothing that loses fine details and nuances in the data. The repeated convolutions may also lead to vanishing gradients, which causes less informative training. However, ChebConv leverages Chebyshev polynomials, which when applied as filters, focus on a small neighborhood around each node that helps preserve the local structure of the graph and mitigate oversmoothing. This is particularly beneficial when dealing with jets that exhibit patterns or structures at various spatial resolutions. ChebConv also adapts well to graphs with different sizes as they operate on the graph structure rather than relying on a fixed-size filter, which becomes advantageous when dealing with jets of different numbers of constituent particles.

\subsubsection{Edge Convolutions for Global Jet Structure Analysis}

The edge convolutional operators are introduced after the Chebyshev convolutional operators in PCN to capture global jet structures. This stage is pivotal for comprehending the overall jet formation and identifying patterns indicative of specific particle origins. EdgeConv excels in capturing the broader relational context within the jet, analyzing how localized interactions aggregate to form distinct jet features observable at larger scales. This global perspective is essential for distinguishing between different types of jets, particularly in complex collision events.

The EdgeConv operation integrates the Chebyshev-processed features:

\begin{align}
\boldsymbol{G'}_{\text{Edge}} = \text{EdgeConv}\left(\boldsymbol{G'}_{\text{Cheb}}, \boldsymbol{W}_{\text{Edge}}^{\ell}\right)
\end{align}

where \(\boldsymbol{W}_{\text{Edge}}^{\ell}\) denotes the learnable weights in the EdgeConv layer.

At its core, EdgeConv utilizes two sets of independently learnable parameters to encode global shape information and local relational information between nodes resulting in its fully expanded form:

\begin{equation} \label{eq:8}
x'_{i} = max_{j\in N(i)}\Theta\cdot (x_{i}-x_{j}) + \Phi \cdot x_{i}
\end{equation}

\noindent where $\Phi$ and $\Theta$ are independently learnable sets of weights.

\subsubsection{Synthesis of Local and Global Features in Jet Tagging}

The architectural decision to interleave EdgeConv layers between ChebConv layers is grounded in the rationale that this configuration facilitates the concurrent extraction of local features through ChebConv and relational information via EdgeConv at each processing stage. While a sequential arrangement of ChebConv layers followed by EdgeConv layers may yield a more hierarchical feature extraction, it defers the integration of global context until the EdgeConv layers. Given the inherently intricate structures exhibited by jets across various resolutions, an interleaved structure is favored for the comprehensive extraction of both local and global features.

Furthermore, the interleaved structure proves advantageous in accommodating different jet compositions, characterized by varying numbers of constituent particles. The adaptability of ChebConv to varying graph sizes is particularly beneficial in this context. The introduction of an EdgeConv layer between ChebConv layers enables the model to dynamically adjust the receptive field, allowing it to flexibly adapt to the diverse structures of jets. In contrast, employing consecutive ChebConv layers may result in the capturing of information solely from the immediate neighborhood of nodes, progressively extending to larger neighborhoods. This approach could potentially lead to the model inadequately capturing the local structures of jets with fewer particles and oversampling those with more particles. An analysis of different model configurations is presented in Section 4.2.

\section{Results}

The performance of PCN and PCN-Lite was evaluated on the JETCLASS dataset introduced by Qu et al. \cite{3}. We investigated evaluations of accuracy, area under the ROC curve (AUC), and area under the precision-recall curve (AUPR) for each model.

\subsection{Dataset and Experiment}

We use the full set of features provided in the JETCLASS dataset and follow the graph construction method detailed in Section 3.1. The generated graphs are given as inputs to our models. The JETCLASS dataset consists of 100M jets for training, 5M for validation, and 20M for testing. Ten different types of jets are provided, evenly distributed into the ten classes. Jet tagging on this dataset is a multi-class classification task, making studies on it more conducive for LHC tasks. Generation of the JETCLASS dataset was through Markov Chain Monte Carlo (MCMC) methods consistent with those used in CERN collaborations. More details about the production of particles and clustering of the jets can be found in Ref. \cite{3}. A sample size of 1M jets for the training process was determined to be sufficient as training with samples of between 2M and 10M provided no increase in performance. The 1M jet events were split 800k/100k/100k for training, validation, and initial testing. Final metrics are derived from the full testing set of 20M jets.

For implementation, we employ the AdamW optimizer \cite{4} with a learning rate of 1e-3. Both models are given a maximum of 500 epochs to train over. A convergence threshold is set to 0.0001, and 10 epochs are given to a model for an improvement in validation loss greater than the convergence threshold. If the improvement is not greater, training is stopped.

PCN and PCN-Lite are evaluated on their accuracy and area under the ROC curve (AUC) to quantify overall performance. We also analyze the area under the precision-recall curves (AUPR) for each model. AUPR is a critical metric for jet tagging to assess how well a model can identify positive instances of a jet without accidentally classifying background instances as positive (false-positive). In addition, the background rejection metric is calculated for each class, as it is directly related to the discovery potential in high-energy physics experiments \cite{3}. Background rejection is only calculated for PCN and follows the methodology described in \cite{3}. Table 1 summarizes the accuracy of PCN and PCN-Lite.

\begin{table}[ht]
\scriptsize
\centering
\begin{tabular}{c c c c c c c c c c c c c} 
 \hline
  & Macro- & \textit{H$\rightarrow$b$\overline{b}$} & \textit{H$\rightarrow$c$\overline{c}$} & \textit{H$\rightarrow$gg} & \textit{H}$\rightarrow$4\textit{q} & \textit{H$\rightarrow$lvqq'} & \textit{t$\rightarrow$bqq'} & \textit{t$\rightarrow$blv} & \textit{W$\rightarrow$qq'} & \textit{Z$\rightarrow$q$\overline{q}$}\\
& Acc & Acc & Acc & Acc & Acc & Acc & Acc & Acc & Acc & Acc \\ 
 \hline
PCN & 0.942 & 0.951 & 0.929 & 0.923 & 0.929 & 0.981 & 0.961 & 0.987 & 0.920 & 0.900\\
PCN-Lite & 0.936 & 0.950 & 0.923 & 0.917 & 0.919 & 0.973 & 0.957 & 0.984 & 0.914 & 0.892\\
 \hline
\end{tabular}
\caption{Accuracy of PCN and PCN-Lite on the 9 signal classes in the JETCLASS dataset.}
\label{table:1}
\end{table}

PCN outperforms PCN-Lite in accuracy across all classes.  A noticeable trend is the consistency of performance between both PCN and PCN-Lite. Both the PCN and PCN-Lite architectures were most accurate in classifying \textit{t$\rightarrow$blv} jets and \textit{H$\rightarrow$lvqq'} jets and were most inaccurate in classifying \textit{Z$\rightarrow$q$\overline{q}$} jets. Table 2 summarizes the AUC of PCN and PCN-Lite.

\begin{table}[h]
\scriptsize
\centering
\begin{tabular}{c c c c c c c c c c c c c} 
 \hline
  & Macro- & \textit{H$\rightarrow$b$\overline{b}$} & \textit{H$\rightarrow$c$\overline{c}$} & \textit{H$\rightarrow$gg} & \textit{H}$\rightarrow$4\textit{q} & \textit{H$\rightarrow$lvqq'} & \textit{t$\rightarrow$bqq'} & \textit{t$\rightarrow$blv} & \textit{W$\rightarrow$qq'} & \textit{Z$\rightarrow$q$\overline{q}$}\\
& AUC & AUC & AUC & AUC & AUC & AUC & AUC & AUC & AUC & AUC \\ 
 \hline
PCN & 0.95 & 0.95 & 0.94 & 0.90 & 1.00 & 0.98 & 0.97 & 0.92 & 0.95 & 0.99 \\
PCN-Lite & 0.94 & 0.94 & 0.93 & 0.88 & 1.00 & 0.98 & 0.97 & 0.90 & 0.94 & 0.99\\
 \hline
\end{tabular}
\caption{AUC of PCN and PCN-Lite on the JETCLASS dataset.}
\label{table:2}
\end{table}

PCN overall outperforms PCN-Lite in AUC, but their performance is equal in some classes. The consistency of performance is present again. Both PCN and PCN-Lite achieve optimal AUC in classifying \textit{H}$\rightarrow$4\textit{q} jets and near-optimal AUC in classifying \textit{Z$\rightarrow$q$\overline{q}$} jets. This indicates the models are strongest in discriminating between instances of these jets compared to others. Both models had their worst AUC's in classifying \textit{H$\rightarrow$gg} jets. 

The observed phenomenon in the vector boson jets, wherein \textit{Z$\rightarrow$q$\overline{q}$} exhibits lower accuracy than \textit{W$\rightarrow$qq'} but yields improved AUC, can be attributed to the inherent differences in the characteristics of the two decay channels. While accuracy focuses on the overall correctness of predictions, AUC considers the model's ability to discriminate between positive and negative instances across a range of decision thresholds. Notably, the model demonstrates a high level of confidence in the predictions it gets right for \textit{W$\rightarrow$qq'} jets, leading to a more concentrated distribution of high-confidence correct predictions. In contrast, the inherent complexity and intricacies of \textit{Z$\rightarrow$q$\overline{q}$} may introduce challenges in achieving consistently high confidence levels for correct predictions, thus impacting overall accuracy. Table 3 summarizes the AUPR of PCN and PCN-Lite.

\begin{table}[ht]
\scriptsize
\centering
\begin{tabular}{c c c c c c c c c c c c c} 
 \hline
  & Macro- & \textit{H$\rightarrow$b$\overline{b}$} & \textit{H$\rightarrow$c$\overline{c}$} & \textit{H$\rightarrow$gg} & \textit{H}$\rightarrow$4\textit{q} & \textit{H$\rightarrow$lvqq'} & \textit{t$\rightarrow$bqq'} & \textit{t$\rightarrow$blv} & \textit{W$\rightarrow$qq'} & \textit{Z$\rightarrow$q$\overline{q}$}\\
& AUPR & AUPR & AUPR & AUPR & AUPR & AUPR & AUPR & AUPR & AUPR & AUPR \\ 
 \hline
PCN & 0.80 & 0.70 & 0.63 & 0.48 & 0.98 & 0.88 & 0.85 & 0.67 & 0.70 & 0.96 \\
PCN-Lite & 0.77 & 0.65 & 0.59 & 0.41 & 0.98 & 0.84 & 0.82 & 0.61 & 0.66 & 0.93\\
 \hline
\end{tabular}
\caption{AUPR of PCN and PCN-Lite on the JETCLASS dataset.}
\label{table:3}
\end{table}

The baselines for the AUPR of each class are equal to the fraction of positives \cite{45}, calculated as the number of positive instances over the total number of instances. In our setup, the test data is near-perfectly balanced, with each class having 2M positive instances. Thus, the baseline AUPR for comparison is 0.1. PCN consistently outperforms PCN-Lite in almost all classes. The performances are also consistent with their respective AUC performances. Both PCN and PCN-Lite achieve the best AUPR in classifying \textit{H}$\rightarrow$4\textit{q} jets and \textit{Z$\rightarrow$q$\overline{q}$} jets and the worst AUPR in classifying \textit{H$\rightarrow$gg} jets. Table 4 summarizes the background rejection of PCN.

\begin{table}[ht]
\scriptsize
\centering
\begin{tabular}{c c c c c c c c c c c c} 
 \hline
    & \textit{H$\rightarrow$b$\overline{b}$} & \textit{H$\rightarrow$c$\overline{c}$} & \textit{H$\rightarrow$gg} & \textit{H}$\rightarrow$4\textit{q} & \textit{H$\rightarrow$lvqq'} & \textit{t$\rightarrow$bqq'} & \textit{t$\rightarrow$blv} & \textit{W$\rightarrow$qq'} & \textit{Z$\rightarrow$q$\overline{q}$}\\
& $Rej_{50\%}$ & $Rej_{50\%}$ & $Rej_{50\%}$ & $Rej_{50\%}$ & $Rej_{99\%}$ & $Rej_{50\%}$ & $Rej_{99.5\%}$ & $Rej_{50\%}$ & $Rej_{50\%}$ \\ 
 \hline
PCN & 15.16 & 13.10 & 25.70 & 253.02 & 32.20 & 27.70 & 12.41 & 78.02 & 1353.02 \\
 \hline
\end{tabular}
\caption{Background rejection of PCN on the JETCLASS dataset.}
\label{table:4}
\end{table}

The background rejection for each signal class (background is considered $q/g$ jets) is relatively consistent with values provided in a previous study \cite{46}. Our discrepancy in values in relation to the previous state-of-the-art, ParT, in Ref. \cite{3} is due to not including the softmax probability function for calculation of the true positive rate (TPR) and false positive rate (FPR). When interpreting our background rejection, PCN approaches event selection by favoring the count of false positives rather than potentially discarding events containing novel or unexpected phenomena. This characteristic enables PCN to capture a greater number of events for further investigation and analysis. 

To corroborate the motivations of interleaving EdgeConv layers between ChebConv layers, we trained and tested three related models:

\begin{enumerate}
  \item PCN-Edge: five layers of EdgeConv with no ChebConv layers
  \item PCN-Inverse: three layers of EdgeConv interleaved with two layers of ChebConv
  \item PCN-Cheb: five layers of ChebConv with no EdgeConv layers
\end{enumerate}

PCN-Edge investigates the model's ability to capture local graph structures and relationships without the additional spectral processing provided by ChebConv layers. We can think of this as extracting only the inter-particle relations and not the particle-level features. PCN-Inverse investigates whether emphasizing local edge-based convolutions over global spectral graph convolutions affects the model's ability to capture essential global features. PCN-Cheb investigates the effect of more ChebConv layers in comparison to PCN-Lite (only three ChebConv layers). Each model was trained on 1M jets with the same 800k/100k/100k for training, validation, and initial testing, and the same final testing set of 20M jets. Implementation follows the same procedure detailed in Section 4.1. The results are summarized in the table below. 

\begin{table}[ht]
\scriptsize
\centering
\begin{tabular}{c c c c c c c c c c c c c} 
 \hline
  & Macro- & \textit{H$\rightarrow$b$\overline{b}$} & \textit{H$\rightarrow$c$\overline{c}$} & \textit{H$\rightarrow$gg} & \textit{H}$\rightarrow$4\textit{q} & \textit{H$\rightarrow$lvqq'} & \textit{t$\rightarrow$bqq'} & \textit{t$\rightarrow$blv} & \textit{W$\rightarrow$qq'} & \textit{Z$\rightarrow$q$\overline{q}$}\\
& Acc & Acc & Acc & Acc & Acc & Acc & Acc & Acc & Acc & Acc \\ 
 \hline 
PCN-Edge & 0.891 & 0.887 & 0.881 & 0.890 & 0.870 & 0.889 & 0.907 & 0.929 & 0.887 & 0.869\\
PCN-Inverse & 0.873 & 0.864 & 0.873 & 0.876 & 0.853 & 0.876 & 0.881 & 0.892 & 0.880 & 0.857\\
PCN-Cheb & 0.853 & 0.822 & 0.858 & 0.856 & 0.839 & 0.860 & 0.856 & 0.876 & 0.810 & 0.869\\
PCN & 0.942 & 0.951 & 0.929 & 0.923 & 0.929 & 0.981 & 0.961 & 0.987 & 0.920 & 0.900\\
PCN-Lite & 0.936 & 0.950 & 0.923 & 0.917 & 0.919 & 0.973 & 0.957 & 0.984 & 0.914 & 0.892\\
 \hline
\end{tabular}
\caption{Accuracies of related PCN models on the 9 signal classes in the JETCLASS dataset.}
\label{table:5}
\end{table}

We observe that for all classes, PCN and PCN-Lite substantially outperform the related models. We can infer two main conclusions from this. First, the performance difference between PCN and the related models supports the interleaved configuration of ChebConv and EdgeConv layers. PCN-Inverse, by prioritizing EdgeConv layers over ChebConv layers, may have become more biased towards local structures, which led to a diminished capability to discern broader graph patterns. Second, the performance difference between PCN-Cheb, PCN-Edge and PCN-Lite suggests that adding more layers, either ChebConv or EdgeConv, does not lead to a more accurate model. In fact, the observed trend prompts the postulation that an excessive number of layers, be it ChebConv or EdgeConv, may constrain the model's capacity to learn the intricacies of the opposite feature type. This supports a balanced layer configuration (like PCN) for optimal performance.

\subsection{Comparison to State-of-the-Art Models}

We compare the performances of PCN and PCN-Lite with four baseline models trained on the JETCLASS dataset: PFN \cite{12}, P-CNN \cite{47}, ParticleNet \cite{5}, and ParT \cite{3}. The Particle Flow Network (PFN) architecture is based on the Deep Sets framework proposed by Zaheer et al. \cite{48}. The P-CNN architecture was used in the CMS experiment by the DeepAK8 algorithm \cite{26}. ParticleNet follows a dynamic graph convolutional neural network similar to PCN using only EdgeConv layers \cite{5}. Particle Transformer (ParT) is the state-of-the-art tagger utilizing a transformer architecture \cite{3}. Table 6 summarizes the comparison of accuracy and AUC with the baseline models.

\begin{table}[ht]
\scriptsize
\centering
\begin{tabular}{c c c} 
 \hline
  & Macro- & Macro-\\
& Accuracy & AUC \\ 
 \hline
PFN & 0.772 & 0.9714\\
P-CNN & 0.809 & 0.9789\\
ParticleNet & 0.844 & 0.9849\\
ParT & 0.861 & \textbf{0.9877}\\
PCN-Lite & 0.936 & 0.9400\\
\textbf{PCN} & \textbf{0.942} & 0.9500\\
 \hline
\end{tabular}
\caption{Comparison of model performances on the JETCLASS dataset.}
\label{table:6}
\end{table}

We conclude that PCN significantly improves upon the state-of-the-art accuracy set by ParT by 8.1\%. PCN-Lite also improves upon ParT and other taggers in accuracy by a notable amount. The improvement in accuracy indicates a reduced number of misclassifications by PCN and PCN-Lite. Both models report lower AUC's than ParT and other taggers, but when contextualized to specific classes, the high AUC in conjunction with improved accuracy leads to a substantial improvement in discriminative power. This is particularly true for the \textit{H}$\rightarrow$4\textit{q} jets and \textit{Z$\rightarrow$q$\overline{q}$} jets, in which PCN achieves near perfect AUC.

\begin{table}[ht]
\scriptsize
\centering
\begin{tabular}{c c c c c} 
 \hline
  & Accuracy & \# params & Time (CPU) [ms]&Time (GPU) [ms]  \\ 
 \hline
PFN & 0.772 & 82k & 0.8 & 0.018\\
P-CNN & 0.809 & 354k & 1.6 & 0.020\\
ParticleNet & 0.844 & 370k & 23 & 0.92\\
PCN-Lite & 0.936 & 148k & 18.6 & 0.024\\
\textbf{PCN} & \textbf{0.942} & 165k & 19.4 & 0.035\\
 \hline
\end{tabular}
\caption{Comparison of number of parameters and inference time on the JETCLASS dataset.}
\label{table:7}
\end{table}

Table 7 demonstrates the model complexity of PCN and PCN-Lite in comparison to previously researched jet tagging models. We observe that the PCN achieves a notable increase in accuracy on the JETCLASS dataset with substantially fewer trainable parameters. This efficiency can be largely attributed to the use of Chebyshev graph convolutions, which enable PCN to capture information from neighboring nodes at different graph distances. Additionally, the spectral nature of Chebyshev polynomials allows them to approximate localized filters more effectively. However, we do note that the inference time for PCN and PCN-Lite is notably longer than that of other models. This can be explained by the increased computational complexity associated with spectral graph convolutions (ChebConv), which involves the computation of eigenvalues and eigenvectors that can lead to longer inference times compared to simpler convolutional methods.

\section{Discussion}

In this work, we propose two performant graph neural networks Particle Chebyshev Network (PCN) and PCN-Lite that utilize a combination of Chebyshev graph convolutions and edge convolutions. PCN achieves state-of-the-art accuracy in classification tasks on the JETCLASS dataset. Future areas of research could extend into the usage of ChebConv in conjunction with other graph convolutional layers for jet tagging applications. Further investigations extending beyond convolution graph operators, such at attention mechanisms could elicit different interactions and performance on JETCLASS.  For more comprehensive comparisons of network performance, fine-tuning for training and testing on the top quark tagging dataset \cite{49} and quark-gluon tagging dataset \cite{50} would be an interesting domain to investigate further.

\end{document}